\documentclass[conference]{IEEEtran}
\IEEEoverridecommandlockouts

\usepackage{cite}
\usepackage{amsmath,amssymb,amsfonts}
\usepackage{algorithmic}
\usepackage{graphicx}
\usepackage{textcomp}
\usepackage{xcolor}
\usepackage{float}
\usepackage[caption=false,font=footnotesize]{subfig}
\usepackage{booktabs}
\usepackage[percent]{overpic}
\usepackage{threeparttable}

\def\BibTeX{{\rm B\kern-.05em{\sc i\kern-.025em b}\kern-.08em
    T\kern-.1667em\lower.7ex\hbox{E}\kern-.125emX}}
\begin{document}

\title{Dual IMU System for Accurate Gastrointestinal Motility Tracking
\\
}


\author{Ziyao Zhou$^{1*}$,~\IEEEmembership{}
Boyan Li$^{1*}$,~\IEEEmembership{}
Chenyang Ren$^{1}$,~\IEEEmembership{}
Tiancheng Cao$^{1}$,~\IEEEmembership{}Dawei Wang$^{1}$,~\IEEEmembership{} Hen-Wei Huang$^{1,2}$~\IEEEmembership{}
        
\thanks{$^{1}$School of Electrical and Electronic Engineering, Nanyang Technological University, Singapore}
\thanks{$^{2}$LKC School of Medicine, Nanyang Technological University, Singapore}
\thanks{$^{*}$Equal Contribution}

}

 \maketitle

\begin{abstract}
Gastrointestinal (GI) motility disorders impair the coordinated movement of food through the digestive tract. The Wireless Motility Capsule was developed to indirectly assess GI motility by recording pressure, pH, and temperature along the GI tract. While the inertial measurement unit (IMU) offers precise direct motion tracking, one IMU system is highly susceptible to artifacts from body movement and respiration. To address this, we propose a dual-IMU system designed to decouple intrinsic GI motility from extrinsic body motion. One IMU is integrated into an ingestible capsule to monitor internal movement, while a second IMU is worn externally to capture whole-body dynamics. Experimental validation was conducted using a turntable to simulate random body movements and a Stewart platform oscillating at 0.1 Hz to emulate GI motility. Three sensor fusion algorithms—Extended Kalman, Madgwick, and Mahony Filter—were benchmarked to compute real-time relative orientation between the two IMUs, represented via quaternion analysis. Results show that the dual-IMU system effectively differentiates platform motion from turntable motion, demonstrating its potential for isolating GI-specific motility. Pearson correlation coefficients exceeding 0.8 between the baseline and noise-cancelled motility signals indicate strong agreement and effective noise suppression.

\end{abstract}


\section{Introduction}
Gastrointestinal (GI) motility is the coordinated, often rhythmic contraction of the GI tract's smooth muscles \cite{grundy1985gastrointestinal}, enabling the propulsion and mixing of luminal contents throughout the digestive system \cite{Sanders2012Regulation}, regulated by multiple neurohumoral factors \cite{ Anandabaskar2021}. Understanding GI motility is fundamental for advancing our knowledge of digestion \cite{Miron2019GASTROINTESTINAL}, identifying root causes of chronic GI disorders \cite{Lacy2006Gastrointestinal}, developing better diagnostic tools and treatments \cite{Poh2011}, and understanding broader physiological systems like the brain-gut axis \cite{Schemann2005Control}.

\begin{figure}[htb]
 \centering
\includegraphics[width=1\linewidth]{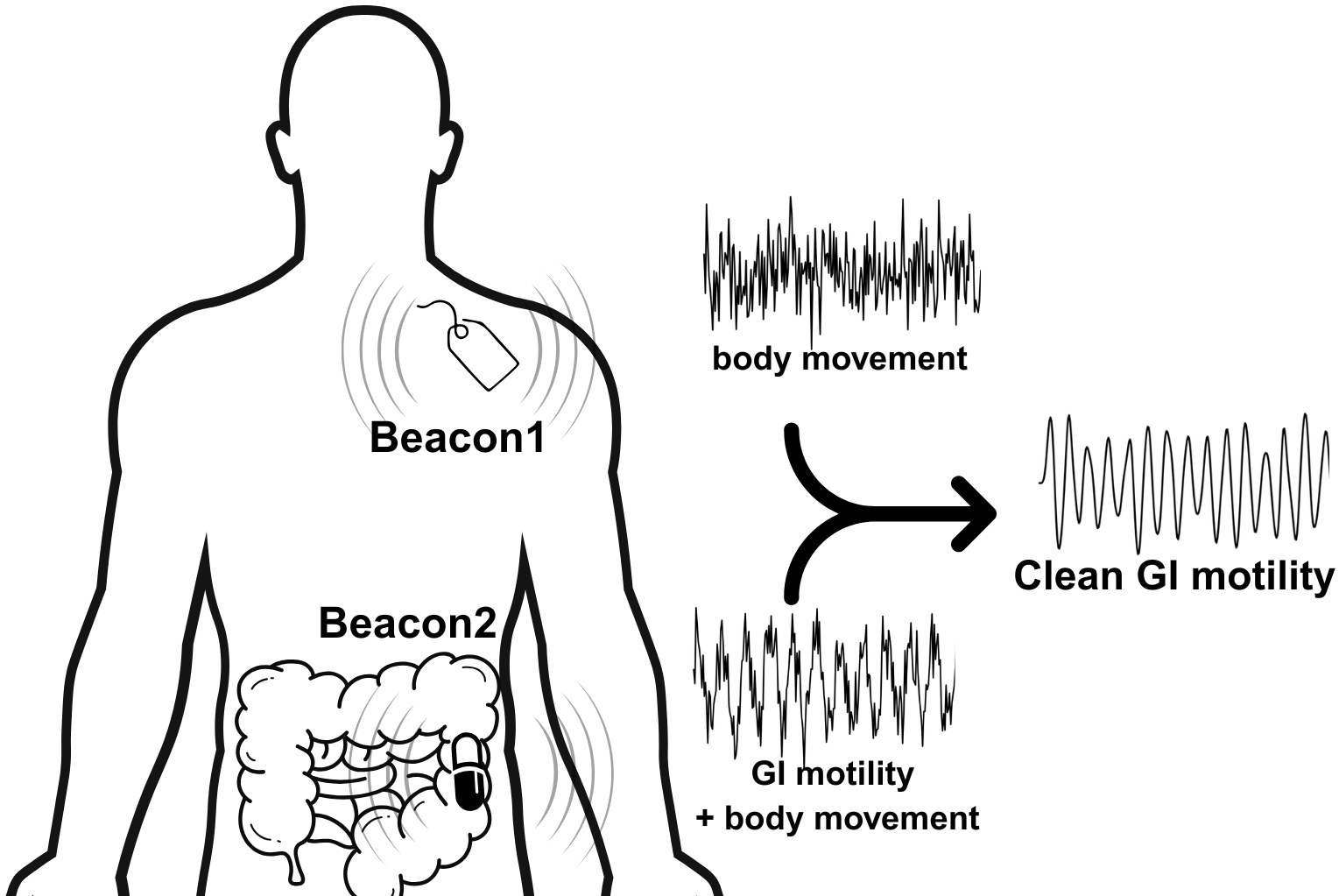}
\caption{Dual IMU system: the system employs two beacons to measure capsule movement in the GI tract and body motion individually.}
  \vspace{-3mm} 
\label{fig_humanbody}

\end{figure}
  
A range of diagnostic approaches have been established to evaluate GI motility. For example, high-resolution manometry for assessing contractile patterns \cite{Gyawali2021A, Mion2018Anal}, radiation studies for GI emptying and transit \cite{Maurer2020Enhancing, Ranjan2013MEASUREMENT}, and electrogastrography for recording myoelectric activity \cite{Kafee2023Electrogastrography}. Each of these techniques requires specialized instrumentation and rigorously controlled testing environments to ensure accuracy and reproducibility.


Another method is the use of the Wireless Motility Capsule (WMC), capable of in situ recording variations in pH, pressure, and temperature in the gut \cite{Camilleri2010, Rao2009, Kuo2008, Maqbool2009}. It indirectly measures GI contractile activity, including contraction frequency, amplitude, and a motility index that are derived from intraluminal pressure patterns \cite{Maqbool2009}. However, intraluminal pressure reflects not only the mechanical force of the gut wall exerted on the sensor, but also the fluid forces within the lumen during contractions\cite{doi:10.1152/ajpgi.00532.2010}, thus compromising the GI motility measurement.

A promising approach for directly capturing GI motion involves integrating an inertial measurement unit (IMU) into an ingestible capsule. While this enables direct motion monitoring of the capsule within the gut, its accuracy is highly sensitive to external body movement, which can introduce significant noise into the measurements \cite{vedaei2021localization}. To address this limitation, we propose a dual-IMU system comprising two beacons. As illustrated in Figure 1, one is integrated into the ingestible capsule (Beacon2) and the other is worn externally on the body (Beacon1). This configuration enables the decoupling of GI-specific motion from whole-body dynamics, allowing for more accurate assessment of intrinsic gut motility.


\section{Methodology}

The system has three main functions, as shown in Figure 2: data capture by IMU beacons, data transmission to the computer by the BLE central devices, attitude estimation and relative motion calculation in Python. 

The design of the beacon is shown in Figure 3(a). 
We designed a compact PCB-based wireless beacon using the nRF52833 MCU and the ICM-20948 IMU. Both IMUs sample at 100 Hz and send data via BLE (Nordic UART Service) to a central device, which forwards it over USB to a computer for analysis. 

A Stewart platform shown in Figure 3(b) is used to simulate the motion in the colon tract by periodically changing the platform's attitude and rotation angle \cite{8333300}. The interval is chosen to be 10 seconds, which is compatible with the 2-6 cycles per minute slow wave in human colon\cite{corsetti2019first}.  

Another rotating turntable shown in Figure 3(c) is used to simulate human body rotation. This rotation could be considered as the noise requiring cancellation. Therefore, we controlled the turntable to randomly alter its rotation direction during operation. 

\begin{figure}[t]
 \centering
\includegraphics[width=1\linewidth]{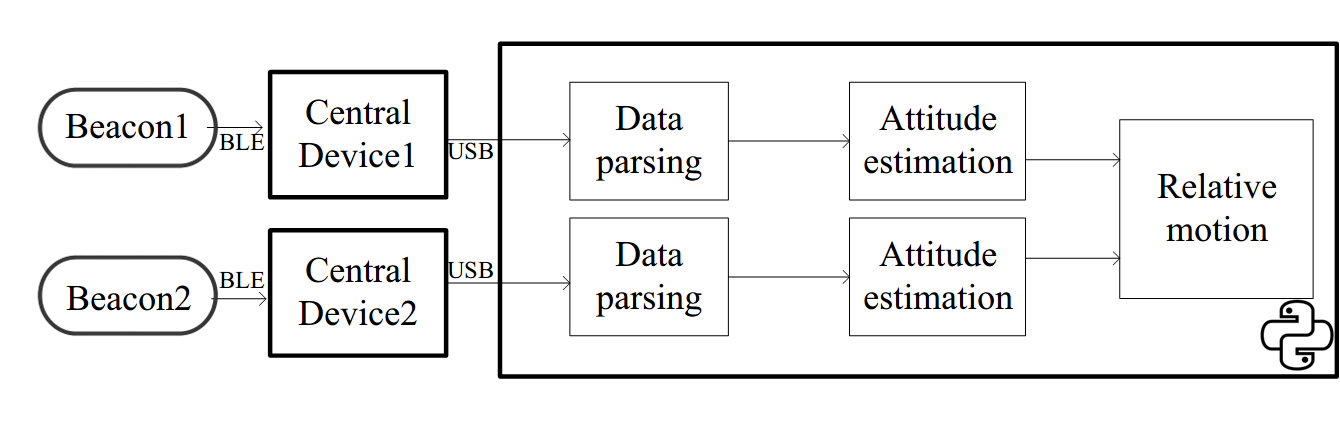}
\caption{Block diagram of dual IMU system. Data from the two beacons is processed in parallel, followed by relative attitude computation.}
\label{fig_diagram}
\end{figure}

\subsection{Attitude estimation}
The raw outputs of the IMU can capture the motion; however, they are referenced only to the sensor's own body-fixed frame \cite{9408659}. By performing attitude estimation, one can derive the object's motion with respect to the earth-fixed frame \cite{5400372}, which facilitates subsequent analyses of relative orientation.

In order to perform attitude estimation, the IMUs in the beacons were first calibrated. All raw IMU data were then smoothed using a low-pass filter to reduce the influence of noise on the final attitude estimation.

We employed and compared three commonly used attitude estimation algorithms: the Extended Kalman, the Madgwick, and the Mahony Filter. The implementations were taken from the Python AHRS library, and all filters were first used with their default parameter settings. For comparison,  clockwise and counterclockwise 90° rotations were performed on each principal axis three times: x for pitch, y for roll, and z for yaw. 

The estimated quaternions were converted to Euler angles and compared to the 90° target under three axes to identify the most accurate one for further fine-tuning. With its parameters properly tuned, the best algorithm processed the beacons' data in parallel. 
\subsection{Relative motion}



When the beacon inside the Gl tract and the external body reference beacon move together, the relative attitude estimation between the two beacons can cancel out the body noise. To derive the relative motion, we compute the relative quaternion $\mathbf{q}_{\text{relative}}$ using two quaternions obtained from two beacons. The calculation is as follows. 

Let $\mathbf{q}_1 = [w_1, x_1, y_1, z_1]$ and $\mathbf{q}_2 = [w_2, x_2, y_2, z_2]$ denote the orientation quaternions of Beacon1 and Beacon2, respectively. The relative rotation is derived by: $\displaystyle \mathbf{q}_{\text{relative}} = \mathbf{q}_2 \otimes \mathbf{q}_1^*$. The relative quaternion is converted into Euler angles for easier analysis. 

\begin{figure}[t]
 \centering
\includegraphics[width=1\linewidth]{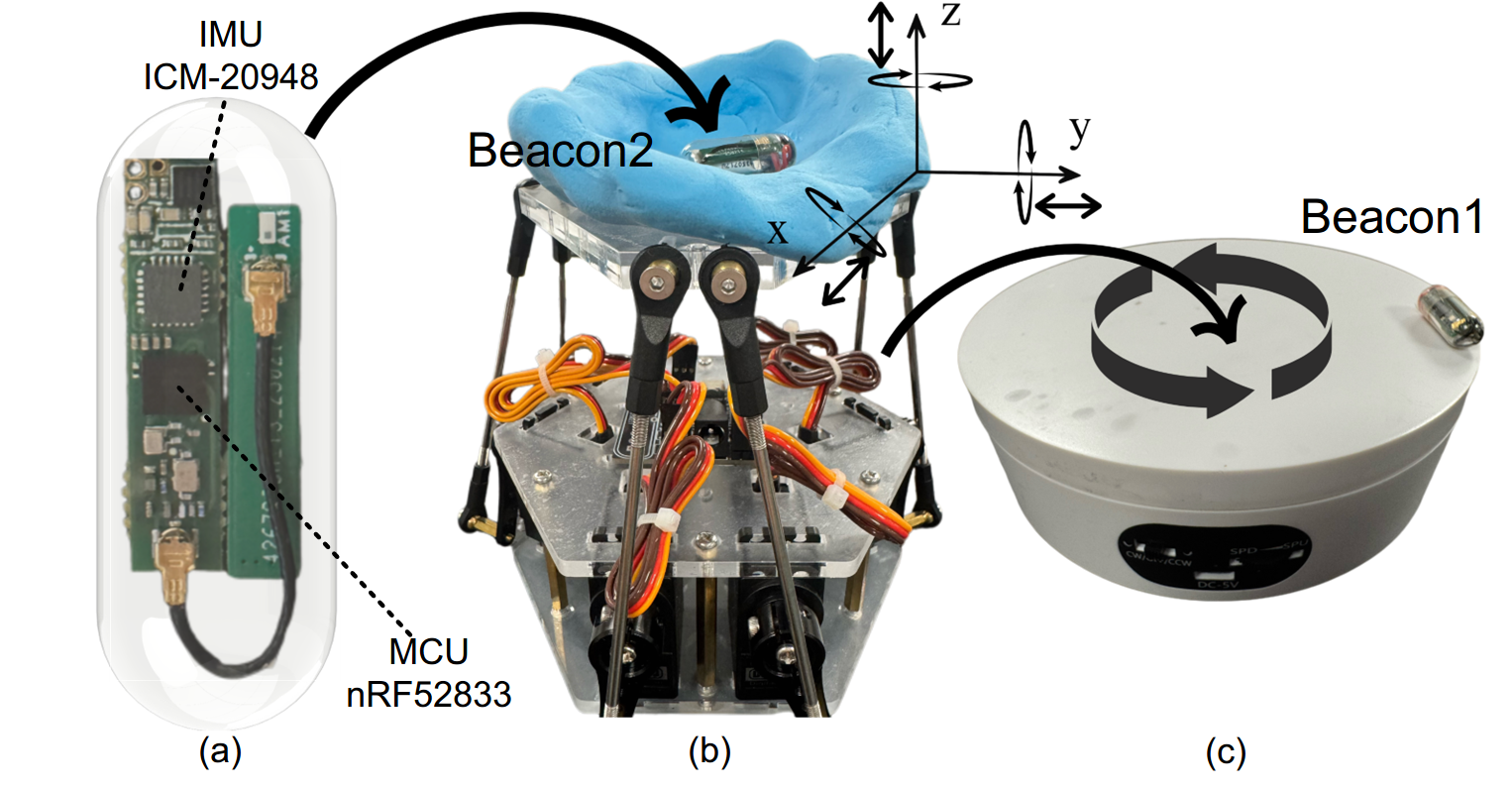}
\caption{(a) Beacon encapsuled in a 000 capsule. (b) The Stewart platform is used to simulate GI motility. (c) The turntable emulates human body noise.}
\label{fig_beacon_platform}
\end{figure}



At first, one beacon was used to capture the simulated motion on the Stewart platform, as shown in Figure 3(b), which served as the baseline. After that, this platform was placed on the turntable, as shown in Figure 3(c). We can obtain the Euler angles of Beacon2's motion before noise cancellation, as well as the final results after noise cancellation. The results are analyzed statistically using Pearson's correlation coefficient to see whether we can get clean simulated GI motility data. 

\section{RESULTS AND DISCUSSION}

\subsection{Attitude estimation}

\begin{figure*}[htbp]
  \centering
  \includegraphics[width=1
  \textwidth]{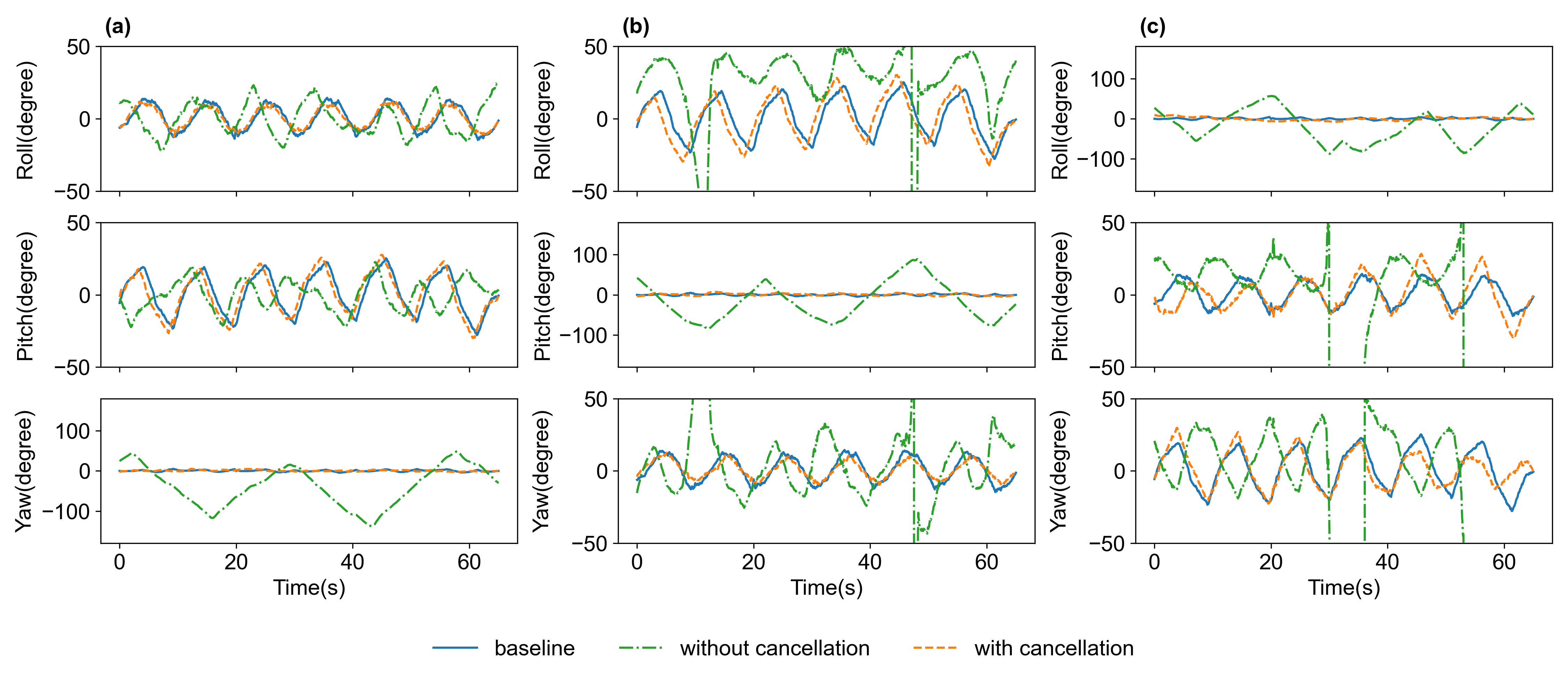}
   \vspace{-5mm} 
  \caption{Comparison of the baseline curve, the curve without body-movement cancellation, and the curve with body-movement cancellation. Panels (a), (b), and (c) show body noise applied on the Beacon2's yaw, pitch, and roll axes, respectively.}
    \vspace{-3mm} 
  \label{Fig_result}
\end{figure*}

After the calibration and filtering, the corrected IMU readings can remain stable and bias-free.

 The comparison of different estimation algorithms is shown in Table~\ref{tab:orientation_comparison}. Extended Kalman algorithm yields mean values of close to 90°. However, all three axes' standard deviations are higher than 2°, suggesting that this filter may not be ideal for stable IMU-based attitude estimation. The Madgwick filter exhibits 75° and a significant deviation of 8.84° in the yaw axis, indicating bad accuracy and stability. 

  \begin{table}[htbp]
\caption{Angle results after three attitudes
estimation algorithms. Avg refers to average, while Std refers to standard deviation.}
\centering
\setlength{\tabcolsep}{3pt}
\begin{tabular}{lcccccc}
\toprule
\textbf{Filter} & \multicolumn{2}{c}{\textbf{Extended Kalman}} & \multicolumn{2}{c}{\textbf{Madgwick}} & \multicolumn{2}{c}{\textbf{Mahony}} \\
\cmidrule(r){2-3} \cmidrule(r){4-5} \cmidrule(r){6-7}
& Avg & Std & Avg & Std & Avg & Std \\
\midrule
Roll angle  & 87.40 & 5.87 & 79.60 & 3.97 & 91.82 & 1.29 \\
Pitch angle & 89.71 & 2.28 & 89.05 & 1.23 & 85.65 & 1.42 \\
Yaw angle   & 83.45 & 4.17 & 75.26 & 8.84 & 85.69 & 2.16 \\
\bottomrule
\end{tabular}

\label{tab:orientation_comparison}
\end{table}

 In contrast, the Mahony filter consistently delivers accurate and stable estimates across all three Euler angles, with both highly accurate means close to 90° and low deviations of 1.29°, 1.42°, and 2.16°. These findings suggest that the Mahony filter is the best algorithm among the three. After tuning the parameters to exhibit the smallest jitter, this parameter set of the Mahony algorithm was applied to the beacon's attitude estimation. 

\subsection{Relative motion}

The real‐time computed Euler angles facilitate analysis of the simulated GI tract's motion frequency, as Figure 4 shows. It presents the results of three relative motion validations. The plots show the Euler angles obtained from left to right when the body movement noise is applied around the Beacon2's yaw (a), pitch(b), and roll axes(c), respectively. 

As shown in Figure~\ref{Fig_result}(a), when the noise happens along the yaw axis of the Beacon2, the yaw angle in the without-cancellation condition (indicated by the green curve) can clearly reflect the noise, which comes from randomly switched turntables. Meanwhile, the pitch and roll angles become highly disordered, making it difficult to interpret the actual motion inside the colon model. In contrast, under the with-cancellation condition (indicated by the orange curve), the rotation along the yaw axis is effectively cancelled out to a flat line that matches the blue baseline. The pitch and roll angles are also aligned with the baseline curve, capturing the periodic motion, which is about 0.1 Hz within the colon model.

Figure~\ref{Fig_result}(b) and (c) show that when the turntable rotates around the pitch and roll axes of Beacon2, the relative attitude estimation successfully cancels out the rotation. Noise on these three principal axes can represent arbitrary spatial noise of human movement.

We quantified the method's effectiveness by computing Pearson correlation coefficients between baseline and noise-removed curves for each axis in three conditions. Results are shown in Table \uppercase\expandafter{\romannumeral 2} .

\begin{table}[htb]
  \centering
  \caption{Pearson correlation coefficient between the baseline and the noise-canceled curve, where values closer to 1 indicate a higher degree of similarity.}
  \label{tab:attitude_angles_swapped}
  \begin{tabular*}{0.6\linewidth}{@{\extracolsep{\fill}} l c c c @{}}
    \toprule
    \textbf{Panels} & \textbf{(a)} & \textbf{(b)} & \textbf{(c)} \\
    \midrule
    \textbf{roll}  & 0.9543 & 0.9149 & 0.0611 \\
    \textbf{pitch} & 0.8042 & 0.2262 & 0.9477 \\
    \textbf{yaw}   & 0.0722 & 0.7347 & 0.8412 \\
    \bottomrule
  \end{tabular*}
    \vspace{-2mm} 
\end{table}

The first column of Table \uppercase\expandafter{\romannumeral 2} indicates the baseline, and the curve with body-movement cancellation exhibits a high degree of similarity when noise is applied to the yaw axis. Correlation analysis revealed that the Pearson coefficients for roll (0.9543) and pitch (0.8042) both exceed 0.8, indicating that the noise-canceled data effectively represent the original motion \cite{HoangShiao2023}. 

However, on the yaw axis, where the body movement is applied, the correlation coefficient remains low (0.0722). This is due to the randomness of noise on the flat line, which eliminates correlations. The other two cases show the same result, indicating that this relative estimation can capture correct simulated GI motility effectively \cite{HoangShiao2023}. 

\section*{Conclusion and Future work}


In this study, we developed a dual-IMU system for accurately tracking GI motility in a dynamic environment. The tested beacon is designed to be ingestible, using an IMU to directly capture movement. After evaluating different attitude estimation algorithms, the Mahony filter demonstrated optimal performance in accuracy and stability. Experiments confirmed that our relative quaternion method effectively captures simulated GI motility of 0.1 Hz by canceling for external body motion. Pearson's correlation coefficients between the baseline and noise‐canceled motility exceeded 0.8, indicating precise real-time monitoring. Future work will consider capsule rolling and perform in vivo tests within actual gastrointestinal environments to verify system performance.

\section{Acknowledgement} \label{thanks}
This work was supported by the Nanyang Professorship, the MOE Tier 1 grant RG71/24, and the MTC MedTech Programmatic Fund M24N9b0125.

\bibliographystyle{ieeetr}

\bibliography{ref} 

@article{Camilleri2010,
  author       = {Camilleri, M. and Thorne, N. K. and Ringel, Y. and Hasler, W. L. and Kuo, B. and Esfandyari, T. and Gupta, A. and Scott, S. M. and McCallum, R. W. and Parkman, H. P. and Soffer, E. and Wilding, G. E. and Semler, J. R. and Rao, S. S.},
  title        = {Wireless pH-motility capsule for colonic transit: prospective comparison with radiopaque markers in chronic constipation},
  journal      = {Neurogastroenterology and Motility},
  year         = {2010},
  volume       = {22},
  number       = {8},
  pages        = {874--e233},
  doi          = {10.1111/j.1365-2982.2010.01517.x},
  publisher    = {Wiley},
}

@article{Rao2009,
  author       = {Rao, Satish S. C. and Camilleri, Michael and Hasler, William L. and Maurer, Alan H. and Parkman, Henry P. and Saad, Richard and Scott, Samantha M. and Semler, Jeffrey R. and Wilding, Gregory E.},
  title        = {Investigation of colonic and whole-gut transit with wireless motility capsule and radiopaque markers in constipation},
  journal      = {Clinical Gastroenterology and Hepatology},
  year         = {2009},
  volume       = {7},
  number       = {5},
  pages        = {537--544},
  doi          = {10.1016/j.cgh.2009.01.017},
  publisher    = {Elsevier},
}

@article{Kuo2008,
  author       = {Kuo, B. and McCallum, R. W. and Koch, K. L. and Sitrin, M. D. and Wo, J. M. and Chey, W. D. and Hasler, W. L. and Lackner, J. M. and Wilding, G. E. and Semler, J. R.},
  title        = {Comparison of gastric emptying of a nondigestible capsule to a radio-labelled meal in healthy and gastroparetic subjects},
  journal      = {Alimentary Pharmacology \& Therapeutics},
  year         = {2008},
  volume       = {27},
  number       = {2},
  pages        = {186--196},
  doi          = {10.1111/j.1365-2036.2007.03564.x},
  publisher    = {Wiley},
}

@article{Maqbool2009,
  author       = {Maqbool, Sabba and Parkman, Henry P. and Friedenberg, Frank K.},
  title        = {Wireless capsule motility: comparison of the SmartPill GI monitoring system with scintigraphy for measuring whole gut transit},
  journal      = {Digestive Diseases and Sciences},
  year         = {2009},
  volume       = {54},
  number       = {10},
  pages        = {2167--2174},
  doi          = {10.1007/s10620-009-0899-9},
  publisher    = {Springer},
}

@book{grundy1985gastrointestinal,
  author    = {Grundy, David},
  title     = {{Gastrointestinal Motility: The Integration of Physiological Mechanisms}},
  publisher = {MTP Press},
  address   = {Lancaster},
  year      = {1985},
}

@article{Sanders2012Regulation,title={Regulation of gastrointestinal motility---insights from smooth muscle biology},author={K. Sanders and S. D. Koh and S. Ro and S. Ward},journal={Nature Reviews Gastroenterology \& Hepatology},year={2012},volume={9},pages={633--645},doi={10.1038/nrgastro.2012.168}}

@incollection{Anandabaskar2021,
  author    = {Anandabaskar, Nishanthi},
  title     = {Drugs Affecting Gastrointestinal Motility},
  booktitle = {Introduction to Basics of Pharmacology and Toxicology: Volume 2: Essentials of Systemic Pharmacology: From Principles to Practice},
  editor    = {Paul, Abialbon and Anandabaskar, Nishanthi and Mathaiyan, Jayanthi and Raj, Gerard Marshall},
  publisher = {Springer Nature Singapore},
  address   = {Singapore},
  year      = {2021},
  pages     = {569--579},
  doi       = {10.1007/978-981-33-6009-9_35},
  isbn      = {978-981-33-6009-9},
  url       = {https://doi.org/10.1007/978-981-33-6009-9_35}
}

@article{Miron2019GASTROINTESTINAL,title={GASTROINTESTINAL MOTILITY DISORDERS IN OBESITY.},author={I. Miron and D. Dumitrascu},journal={Acta endocrinologica},year={2019},volume={15 4},pages={ 497-504 },doi={10.4183/aeb.2019.497}}

@article{Schemann2005Control,title={Control of gastrointestinal motility by the "gut brain"--the enteric nervous system.},author={M. Schemann},journal={Journal of pediatric gastroenterology and nutrition},year={2005},volume={41 Suppl 1},pages={ S4-6 },doi={10.1097/01.SCS.0000180285.51365.55}}

@article{Lacy2006Gastrointestinal,title={Gastrointestinal Motility Disorders: An Update},author={Brian E. Lacy and K. Weiser},journal={Digestive Diseases},year={2006},volume={24},pages={228 - 242},doi={10.1159/000092876}}

@article{Gyawali2021A,title={A Short History of High-Resolution Esophageal Manometry},author={C. Gyawali and P. Kahrilas},journal={Dysphagia},year={2021},volume={38},pages={586-595},doi={10.1007/s00455-021-10372-7}}

@article{Mion2018Anal,title={Anal sphincter function as assessed by 3D high definition anorectal manometry.},author={F. Mion and A. Garros and F. Subtil and H. Damon and S. Roman},journal={Clinics and research in hepatology and gastroenterology},year={2018},volume={42 4},pages={ 378-381 },doi={10.1016/j.clinre.2017.12.004}}

@article{Maurer2020Enhancing,title={Enhancing Scintigraphy for Evaluation of Gastric, Small Bowel, and Colonic Motility.},author={A. Maurer},journal={Gastroenterology clinics of North America},year={2020},volume={49 3},pages={ 499-517 },doi={10.1016/j.gtc.2020.04.006}}

@misc{Ranjan2013MEASUREMENT,
  author = {Ranjan, P.},
  title  = {Measurement of Colonic Transit},
  year   = {2013},
  note   = {Unpublished/undated source (journal not specified).}
}

@article{Kafee2023Electrogastrography,title={Electrogastrography in patients with gastric motility disorders},author={Abdullah Al Kafee and Yusuf Kayar},journal={Proceedings of the Institution of Mechanical Engineers, Part H: Journal of Engineering in Medicine},year={2023},volume={238},pages={22 - 32},doi={10.1177/09544119231212269}}

@article{doi:10.1152/ajpgi.00532.2010,
author = {Dinning, P. G. and Arkwright, J. W. and Costa, M. and Wiklendt, L. and Hennig, G. and Brookes, S. J. H. and Spencer, N. J.},
title = {Temporal relationships between wall motion, intraluminal pressure, and flow in the isolated rabbit small intestine},
journal = {American Journal of Physiology-Gastrointestinal and Liver Physiology},
volume = {300},
number = {4},
pages = {G577-G585},
year = {2011},
doi = {10.1152/ajpgi.00532.2010},
    note ={PMID: 21193528},

}

@article{corsetti2019first,
  title={First translational consensus on terminology and definitions of colonic motility in animals and humans studied by manometric and other techniques},
  author={Corsetti, Maura and Costa, Marcello and Bassotti, Gabrio and Bharucha, Adil E and Borrelli, Osvaldo and Dinning, Phil and Di Lorenzo, Carlo and Huizinga, Jan D and Jimenez, Marcel and Rao, Satish and others},
  journal={Nature Reviews Gastroenterology \& Hepatology},
  volume={16},
  number={9},
  pages={559--579},
  year={2019},
  publisher={Nature Publishing Group UK London}
}

@INPROCEEDINGS{5400372,
  author={Bonnable, Silv{\`e}re and Martin, Philippe and Sala{\"u}n, Erwan},
  booktitle={Proceedings of the 48th IEEE Conference on Decision and Control (CDC) held jointly with 2009 28th Chinese Control Conference},
  title={Invariant Extended Kalman Filter: theory and application to a velocity-aided attitude estimation problem},
  year={2009},
  volume={},
  number={},
  pages={1297-1304},
  doi={10.1109/CDC.2009.5400372}}

@ARTICLE{9408659,
  author={Kecsk{\'e}s, Istv{\'a}n and Odry, {\'A}kos and Tadi{\'c}, Vladimir and Odry, P{\'e}ter},
  journal={IEEE Sensors Journal},
  title={Simultaneous Calibration of a Hexapod Robot and an IMU Sensor Model Based on Raw Measurements},
  year={2021},
  volume={21},
  number={13},
  pages={14887--14898},
  doi={10.1109/JSEN.2021.3074272}}

@INPROCEEDINGS{8333300,
  author={Patel, Vatsal and Krishnan, Sanjay and Goncalves, Aimee and Goldberg, Ken},
  booktitle={2018 International Symposium on Medical Robotics (ISMR)}, 
  title={SPRK: A low-cost stewart platform for motion study in surgical robotics}, 
  year={2018},
  volume={},
  number={},
  pages={1-6},
  doi={10.1109/ISMR.2018.8333300}}

@article{HoangShiao2023,
  author    = {T. Hoang and Y. Shiao},
  title     = {New Method for Reduced-Number IMU Estimation in Observing Human Joint Motion},
  journal   = {Sensors (Basel)},
  year      = {2023},
  month     = jun,
  volume    = {23},
  number    = {12},
  pages     = {5712},
  doi       = {10.3390/s23125712},
  pmid      = {37420876},
  pmcid     = {PMC10302940}
}

@inproceedings{Poh2011,
  author    = {Poh, Y. C. and Buist, M. L.},
  title     = {A computational approach to understanding gastrointestinal motility in health and disease},
  booktitle = {Proceedings of the Annual International Conference of the IEEE Engineering in Medicine and Biology Society},
  year      = {2011},
  pages     = {429--432},
  doi       = {10.1109/IEMBS.2011.6090057},
}

@article{vedaei2021localization,
  author  = {Vedaei, S. S. and Wahid, K. A.},
  title   = {A localization method for wireless capsule endoscopy using side wall cameras and {IMU} sensor},
  journal = {Scientific Reports},
  year    = {2021},
  volume  = {11},
  number  = {1},
  pages   = {11204},
  month   = may,
  doi     = {10.1038/s41598-021-90523-w},
  url     = {https://doi.org/10.1038/s41598-021-90523-w}
}

\end{document}